\documentclass[aps,twocolumn]{revtex4}

\newcommand{\rt}{\mathrm{\scriptscriptstyle T}}
\newcommand{\rf}{\mathrm{\scriptscriptstyle F}}
\newcommand{\phit}{\phi_\rt}
\newcommand{\phif}{\phi_\rf}
\newcommand{\Ut}{U_\rt}
\newcommand{\Uf}{U_\rf}
\newcommand{\M}{\mathrm{\scriptscriptstyle M}}
\newcommand{\be}{\begin{equation}}
\newcommand{\ee}{\end{equation}}
\newcommand{\bea}{\begin{eqnarray}}
\newcommand{\eea}{\end{eqnarray}}
\newcommand{\SI}{\mathcal{S}}
\newcommand{\eqref}[1]{(\ref{#1})}

\begin{document}
\title{Vacuum Decay on a Brane World}

\author{Stephen C. Davis and  Sylvain Br\'echet}
\affiliation{Institute of Theoretical Physics, \'Ecole Polytechnique
F\'ed\'erale de Lausanne, CH--1015 Lausanne, Switzerland}

\begin{abstract}
The bubble nucleation rate for a first order phase transition occurring
on a brane world is calculated. Both the Coleman-de Luccia thin wall
instanton and the Hawking-Moss instanton are considered. The results
are compared with the corresponding nucleation rates for standard
four-dimensional gravity.
\end{abstract}

\maketitle


\section{Introduction}

It is widely believed that as the universe cools it undergoes a series
of phase transitions. Each of these involves a transition from a
metastable ``false vacuum'' ground state to a stable one, the ``true
vacuum''. In the case of a first order  transition, this change is
initiated by the nucleation of bubbles of true vacuum. These then
expand at a speed asymptotically approaching that of light, converting
false vacuum into true as they grow. The decay of false vacuum was
first studied by Kobzarev, Okun and Voloshin~\cite{KOV}. It was shown
by Coleman~\cite{CC} that the nucleation rate could be calculated
using instantons. This was extended by Coleman and de Luccia, who
showed that  the rate at which the decay occurs is altered by
gravity~\cite{CdL}. This will be most significant in the early
universe when the curvature is high. Most previous work on the effects of
gravity on vacuum decay has assumed, not surprisingly, that gravity is
described by general relativity. However this may not be true, in
which case the  phase transition, and any related processes such as
defect formation, will be altered.

There has been a lot of interest in the string-motivated brane world
scenario, in which our universe is a 3-brane embedded in a higher
dimensional ``bulk'' space-time. The extra dimensional effects produce
modifications to gravity. The particular brane model that we will consider
is the Randall-Sundrum II model~\cite{RS}, which has a single extra
dimension and a warped bulk spacetime. The warping allows
conventional gravity to be obtained on the brane in the small
curvature limit. However in more extreme circumstances, such as during
the early universe~\cite{BWcos}, gravity will behave differently.

In this article we calculate the vacuum decay rate on a brane
world. In section~\ref{sec:B} we extend the work of Coleman and de
Luccia (CdL) to the Randall-Sundrum model. Analytic determination of
the correct instanton for this method is not possible, and so we will
instead use some approximate solutions. In section~\ref{sec:TW}  we
look at the CdL thin wall instanton, and consider some limiting cases
of it. In section~\ref{sec:HM} we find the decay rate for the
Hawking-Moss instanton~\cite{HM}. Although we are unable to provide a
completely general analysis, our results do allow vacuum decay in the
brane world to be compared with the results for standard gravity.

\section{Bounce Action}
\label{sec:B}

We will consider a toy model with one scalar field, $\phi$, whose 
potential is $U(\phi)$. The potential has minimums at $\phif$ and
$\phit$, and $\Uf \equiv U(\phif)$ is greater than $\Ut \equiv U(\phit)$. Thus
$\phi=\phif$ is a metastable false vacuum state, and $\phi=\phit$ is
the global minimum of the potential and hence the true vacuum
state. In this article we will take $\Uf$ and $\Ut$ to be both positive.

In the semi-classical limit, the bubble nucleation rate per unit
volume is given by~\cite{CC}
\be
\Gamma/V = A e^{-B/\hbar}[1 + O(\hbar)] \ .
\label{Gamma}
\ee
The bounce action, $B$, is defined as 
\be
B=S_E(\phi)-S_E(\phif) 
\ee
where $S_E$ is the Euclidean action, defined as minus the analytic
continuation of the usual action to imaginary time. The instanton
$\phi$, called the ``bounce'', is the solution of the field equations
which minimises $B$ and which traverses the potential barrier between
the true and false vacuums. In this article we will ignore the factor
$A$, and just determine $B$, since it gives the dominant
contribution to the behaviour of $\Gamma/V$.

For a brane world, with a $Z_2$ symmetric bulk, the Euclidean action
is given by $S_E=S_E^{(B)}+S_E^{(b)}$, where the bulk and brane
contributions are respectively
\be
S_E^{(B)}=\frac{1}{2\kappa_5^2}\int d^5x\sqrt{g}
\left(-R^{(5)}+2\Lambda_5\right) \ ,
\ee
\bea
&&S_E^{(b)}= \frac{1}{\kappa_5^2}\int d^4x\sqrt{h}
\left(2K+\lambda_4\right) \nonumber \\ && \hspace{0.7in}
+ \int d^4x\sqrt{h}\left(\frac{1}{2}(D\phi)^2+U(\phi)\right)\ .
\eea
The induced metric metric on the brane is $h_{ab}=g_{ab}-n_a n_b$,
where $n_a$ is the (outward) brane normal. $K=h^{ab}K_{ab}$ is the
Gibbons-Hawking boundary term~\cite{GH}, with $K_{ab}= h_a^c\nabla_c n_b$ 
being the extrinsic curvature. Although we have a five
dimensional gravity theory, the scalar field is restricted
to the brane. The induced derivative is defined as
$D_a\phi=h_a^b\nabla_b\phi$. 

As with the Randall-Sundrum model, we take
$\Lambda_5 = -6/\ell^2$ and $\lambda_4 = 6/\ell$. We could have absorbed
$\lambda_4$ into $U$, although by keeping them separate we can more
easily compare our results with those of standard four-dimensional gravity.
The effective four dimensional gravitational coupling on the brane is given by
$\kappa^2=\kappa_5^2/\ell$.

Variation of the action gives the field equations
\be
G^{(5)}_{ab} + \Lambda_5 g_{ab} = 0
\label{Beq}
\ee
\bea
-2\left(K_{ab} - K h_{ab}\right) + \lambda_4 h_{ab} 
\hspace{1.3in} &&\nonumber \\ 
= \kappa_5^2\left(D_a\phi D_b\phi
-\frac{1}{2}h_{ab}(D\phi)^2-h_{ab}U(\phi)\right)
&& \label{beq}
\eea
\be
D^2\phi=\frac{dU}{d\phi} \ .
\label{Peq}
\ee

In flat space the bounce is $O(4)$ symmetric~\cite{flatO4}. This is
usually assumed when Einstein gravity is included in the theory, and
we will also assume it for the brane world. This implies
\be
ds^2_\mathrm{brane} = d\xi^2+a(\xi)^2d\Omega_3^2 \ ,  \ \ \ \phi =\phi(\xi)
\ , \label{bmetric}
\ee
where $d\Omega_3^2$ is the metric of $S^3$. For the solution to have a
finite action, we require $a(0) = a(\xi_\mathrm{max}) = 0$ and 
$\dot \phi(0) = \dot \phi(\xi_\mathrm{max}) = 0$.

The general $O(4)$ symmetric solution of the field
equations~(\ref{Beq}--\ref{Peq}) can be obtained be analytically
continuing a brane world cosmology solution. Using the results of
Ida~\cite{Ida}, we obtain 
\be
ds^2_\mathrm{Bulk}=f(r)d\Xi^2+\frac{dr^2}{f(r)}+r^2d\Omega_3^2 \ ,
\label{Bmetric}
\ee
where $f(r)=1+r^2/\ell^2-\mathcal{C}/r^2$. The parameter $\mathcal{C}$
corresponds to the black hole mass in the original cosmological
solution. A brane-based observer would perceive it as ``dark radiation''.
The position of the brane is given by $r=a(\xi)$,
$\Xi=\Xi_b(\xi)$. In each half of the $Z_2$ symmetric bulk,
$r$ ranges from 0 to $a$.

The brane field equations~\eqref{beq} and \eqref{Peq} reduce to
\be
\frac{\dot a^2}{a^2} = \frac{f(a)}{a^2} - \left\{\frac{\kappa_5^2}{6}
\left(U -\frac{\dot \phi^2}{2}\right) + \frac{1}{\ell} \right\}^2 \ ,
\label{Friedmann}
\ee
\be
\ddot \phi
+\frac{3\dot a}{a}\dot \phi-\frac{dU(\phi)}{d\phi}=0 \ .
\label{phieq} 
\ee

The fact that the two metrics agree at $r=a$, gives us the relation
\be
\frac{d\Xi}{d\xi} = \frac{\sqrt{f(a)-\dot a^2}}{f(a)}
\label{coords}
\ee
on the brane.

With the aid of the field equation~\eqref{Beq}, the bulk part of the action
simplifies to
\be
S_E^{(B)} = -\frac{8\pi^2}{3\kappa_5^2}\Lambda_5 
\int_{r=0}^{r=a} r^3 \, dr \, d\Xi  \ . 
\ee
Integrating with respect to $r$, and using the relation~\eqref{coords},
we obtain
\be
S_E^{(B)}=\frac{2\pi^2}{3\ell^2} \int d\xi \frac{a^5}{f(a)}
\left(U -\frac{\dot \phi^2}{2} + \frac{6}{\kappa_5^2\ell}\right) \ .
\label{SB}
\ee

Using the trace of the junction condition~\eqref{beq}, the brane part
of the action reduces to
\be
S_E^{(b)}=-\frac{2\pi^2}{3}\int d\xi a^3
\left(U-\frac{\dot \phi^2}{2}+\frac{6}{\kappa_5^2\ell}\right) \ .
\label{Sb}
\ee

Before we try to evaluate the expression for $B=S_E(\phi)-S_E(\phif)$,
it is convenient to add a total derivative of the form 
$-(2/\kappa^2)\partial_{\xi}(a\dot a^2)$ to the action. This
will not alter the value of $B$, although it does make the thin wall
approximation (see next section) easier to apply. Adding together all
the above contributions to the action, we obtain 
\bea
&&S_E(\phi) = 2\pi^2\int d\xi a^3 \Biggl\{2 U
\nonumber \\ && \hspace{0.3in} {}
+\frac{1}{3} \left(U-\frac{\dot \phi^2}{2}\right)
\left[\frac{a^2}{\ell^2f(a)}-1 
+\frac{\kappa_5^2\ell}{2} \left(\frac{\dot \phi^2}{2}+U\right) \right]
\nonumber \\ && \hspace{0.3in} {}
+ \frac{2}{\kappa_5^2\ell}
\left(\frac{a^2}{\ell^2 f(a)} - \frac{\ell^2}{a^2}[f(a)+1]\right)
\Biggr\} \ .
\label{SE1}
\eea

\section{Thin Wall Approximation}
\label{sec:TW}

The expression for $B=S_E(\phi)-S_E(\phif)$ is too complicated to
evaluate analytically, so we will instead use the thin wall
approximation. This consists of taking the bounce instanton to be a
ball of true vacuum surrounded by false vacuum, with a thin
wall at $a=\bar a$ separating the two regions. Away from the wall
$\phi$ is constant and so the action simplifies considerably. The
thin wall approximation holds when the wall thickness is far smaller
than $\bar a$, which will be the case when the energy difference
$\epsilon=\Uf-\Ut$ is small compared to the barrier height.

On the wall $a$ is roughly constant, and so in the scalar field
equation~\eqref{phieq} the second term can be dropped. We also
approximate the potential on the wall by 
$U_0(\phi) = U(\phi)+O(\epsilon)$. The approximate potential $U_0$
also has minimums at $\phit$ and $\phif$, but with
$U_0(\phif)=U_0(\phit)$. To leading order in $\epsilon$, we can
approximate the equation for $\phi$ on the wall by $\ddot
\phi=dU_0/d\phi$. This is solved by
\be
\frac{1}{2}(\dot \phi)^2-U_0(\phi)=-U_0(\phif) \ ,
\label{approxphi}
\ee
which allows the leading order contribution to $B$ from the wall to be
obtained.

We will now use the thin wall approximation for the brane
action~\eqref{SE1}, and evaluate the parameter $B$.
For simplicity we will assume there is no black hole in the bulk
space, so $\mathcal{C}=0$. For the space outside the wall we have 
$B_\mathrm{outside}=S_E(\phif)-S_E(\phif)=0$. On the wall we use the
expression~\eqref{approxphi} and $a = \bar a$, to obtain
\be
B_\mathrm{wall}=S_E(\phi)-S_E(\phif)= 2\pi^2\bar{a}^3 \alpha_0 S_1\ ,
\ee
where $\alpha_0=(1+\kappa_5^2\ell U_0(\phif)/6)$ and
\be
S_1 
=\int_{\phit}^{\phif}d\phi\{2[U_0(\phi)-U_0(\phif)]\}^{1/2} \ .
\ee

Before evaluating the contribution to $B$ from inside the wall, we
will define some notation
\be
\alpha_i=1+\frac{\kappa_5^2\ell U_i}{6}\ ,
\ee
\be
H^2_i=\frac{\kappa_5^4}{36}U_i^2+\frac{\kappa_5^2}{3\ell}U_i 
= \frac{\alpha_i^2-1}{\ell^2}\ .
\ee
with $i$ being `T' or `F'.

Inside the wall, $\phi$ is a constant. The Friedmann
equation~\eqref{Friedmann} implies
\be
d\xi= \pm da\left(1-H^2_\rt a^2\right)^{-1/2} \ .
\label{xi-a}
\ee
The choice of sign depends which half of the four dimensional
spacetime we are in. The positive sign applies when $\xi$ is near 0,
and the negative sign when it is near $\xi_\mathrm{max}$. We see that
the above coordinate change is not one to
one and that, for a given $\bar{a}$, there are two possible CdL instantons.

If the upper sign in eq.~\eqref{xi-a} applies for the entire region inside
the wall then, after changing variables, we obtain $B_\mathrm{inside} =
\SI(\phit,\bar a) - \SI(\phif,\bar a)$, where
\bea
&&\SI(\phi_i, b)=\frac{4\pi^2}{\kappa^2}\int_{0}^{b} 
\frac{ada}{(1-H^2_i a^2)^{1/2}}
\nonumber \\ && \hspace{0.7in} \times
\left\{1-\frac{a^2}{\ell^2+a^2}\alpha_i  -3(1-H^2_i a^2)
\right\}\ , \hspace{0.3in}
\eea
i.e.\ the contribution to $S_E$ (for constant $\phi$) from the
region $0 < a < b$. The above expression evaluates to
\bea
\SI(\phi_i, a)=\frac{4\pi^2}{\kappa^2} \Biggl\{
\frac{\ell^2}{2} \ln \frac{(\sqrt{1-H^2_ia^2}-\alpha_i)
(1+\alpha_i)}{(\sqrt{1-H^2_ia^2}+\alpha_i)(1-\alpha_i)}
\hspace{0.29in} && \nonumber\\ {}
+ \frac{\ell^2}{\alpha_i+1} (1-H^2_ia^2)^{1/2}
+\frac{1}{H^2_i}(1-H^2_ia^2)^{3/2} -\frac{\alpha_i}{H^2_i}
\! \Biggr\} \ \ &&
\eea
which will give $B_\mathrm{inside}$. 

It is also possible to have a thin wall instanton which has different
signs in the relation~\eqref{xi-a} on different sides of the wall. In
this case the part of $S_E(\phif)$ inside the wall will have
contributions from both halves of the spacetime, and we find
\be
B_\mathrm{inside}=\SI(\phit,\bar a)-2\SI(\phif,1/H_\rf) +\SI(\phif,\bar a) 
\label{Bin2}
\ee
with $1/H_\rf$ being the maximum value of $a$ for $\phi \equiv \phif$.

The expression~\eqref{Gamma} for $\Gamma/V$ is evaluated for the instanton
which minimises the action $B$. We can estimate this by minimising the
our approximate expression for $B$ with respect to $\bar a$. Using
the above expressions, we find that this is the case when
\bea
&&\frac{3}{2}\kappa^2 \bar{a} S_1 \alpha_0 + 
\biggl\{\left[1-\frac{\bar a^2}{\ell^2+\bar a^2}\alpha_\rt\right]
(1-H^2_\rt\bar{a}^2)^{-1/2} 
\nonumber \\ && \hspace{0.4in} {}
- 3(1-H^2_\rt \bar a^2)^{1/2}\biggr\}
- \sigma \biggl\{\rt \leftrightarrow \rf \biggr\} = 0 \ .
\label{aeq}
\eea
The sign $\sigma$ is equal to $-1$ if the expression~\eqref{Bin2} is
applicable, and 1 otherwise.

At late times brane cosmology will reduce to standard four dimensional
cosmology. This occurs when $\ell$ is small relative to other length
scales in the theory. Similarly we expect to obtain the usual four
dimensional tunnelling rate when $\ell$ is small.
Taking $\ell \ll  6/(\kappa_5^2 U)$ and $\ell \ll \bar a$, the
equations for $B$ and $\bar a$ reduce to the
conventional expressions, as obtained by Parke~\cite{Parke}.

Even with the simplifications resulting from the thin wall
approximation, eq.~\eqref{aeq} is still algebraically complicated. Rather than
trying to solve it analytically, it is more instructive to look at various
limiting cases. Gravitational effects will be most significant when
the vacuum energy or the barrier size is large, and so we will
concentrate on these limiting cases.

\subsection*{(i) Large Vacuum Energy Limit}

The expression for $\bar a$ will simplify if we suppose that
$\Uf, \Ut \gg \epsilon$. In this case $\bar a$
will be close to its maximal value, and so $\bar a^2 = (1-\eta)/H_\rf^2$
with $\epsilon/U \ll \eta \ll 1$. We take $\sigma=1$ in eq.~\eqref{aeq}.

If $\kappa_5^2 \ell \Uf \ll \eta$, then higher dimensional
contributions to gravity will be small. In this case we obtain the
same results as for standard four-dimensional gravity. These are
$\eta \sim \epsilon^2/(\kappa^2 U S_1^2)$ and
\be
B_{4D} = \frac{6\sqrt{3}\pi^2S_1}{\kappa^3 U^{3/2}} \ .
\ee
The limits used will be self consistent if $\epsilon/U \ll S_1^2
\kappa^2/\epsilon \ll 1$ and  $\ell \ll \epsilon/(S_1\kappa^2 U)$. 
This case is contained within the limit (i) in ref.~\cite{Parke}.
For comparison, the bounce action in the absence of gravity is
$B_0=27\pi^2 S_1^4/(2\epsilon^3)$, and so we see that the effects of
gravitation decrease $B$.

For higher vacuum energies brane gravity effects will be
important. They will be most significant when
$\kappa_5^2\ell U\gg 1$. In this limiting brane gravity case
$\eta^{3/2} \sim \epsilon/(\kappa_5^2 U S_1)$ and
\be
B_{BW} = \frac{72\pi^2 \ell S_1}{\kappa^3\kappa_5\ell^{1/2} U^2} \ll B_{4D} \ .
\ee
Consistency requires $\epsilon/[\kappa_5^2 \ell U^2] \ll 
S_1^2 \kappa^2/\epsilon \ll U/[\kappa_5^2 \ell \epsilon^2]$.

For both the above expressions, it is the contribution from the wall
that dominates $B$. The inclusion of gravitation in the theory
reduces $B$, and so increases the nucleation rate. If we use brane
gravity instead of conventional gravity, the nucleation
rate is increased even further, especially when $U$ is greater than the
brane tension, $\lambda_4/\kappa_5^2$.

\subsection*{(ii) Large Barrier Limit}

We can also obtain approximate analytic expressions for $\bar
a$ and $B$ when $H_\rf \bar a \ll 1$ and $\sigma=-1$. This implies that
the bounce instanton is close to the maximum size that will fit in the
spacetime. This case corresponds to the limit (ii) considered in
ref.~\cite{Parke}.

For $\kappa_5^2 \ell \Uf \ll 1$ we re-obtain the result for
conventional gravity. This has $\bar a \sim
1/(\kappa^2S_1^2)$ and, using eq.~\eqref{Bin2}, 
\be
B_{4D} = \frac{24\pi^2}{\kappa^4\Uf} \ .
\ee
This limit is valid when $\kappa^2 S_1^2 \gg \Uf$, in other words
when the barrier height is large compared to the vacuum energy.

Brane effects will be most significant when $\kappa_5^2 \ell \Uf \gg
1$. In this case we find $\bar a \sim \kappa_5^2 \ell S_1$ and
\be
B_{BW} = \frac{1152\pi^2}{\kappa^4 \kappa_5^4 \ell^2 \Uf^3} \ll B_{4D}
\ee
which is valid when $\kappa_5^4 S_1^2 \gg 1/(\kappa_5^2\ell \Uf)$.

For this limiting case the dominant contribution to $B$ comes from the
region inside the wall. Again we wee that gravity increases the
nucleation rate, and brane gravity increases it even more.

\section{Hawking-Moss instanton}
\label{sec:HM}

As well as the approximate CdL instanton, we will also consider the
Hawking-Moss instanton. For this $\phi$ is a constant, and sits at the
top of the potential barrier, which we denote by $\phi=\phi_\M$.  If
$U(\phi)$ is very large, it is expected that the Hawking-Moss instanton will
provide a better approximation of bounce than the thin wall CdL
instanton.  The bounce action is
$B=2\SI(\phi_\M,1/H_\M)-2\SI(\phif,1/H_\rf)$, which evaluates to
\be
B= \frac{4\pi^2}{\kappa^2}\left(\frac{\ell^2}{2}
\ln\frac{(\alpha_\M +1)(\alpha_\rf -1)}{(\alpha_\M -1)(\alpha_\rf +1)}
-\frac{\alpha_\M}{H^2_\M}+ \frac{\alpha_\rf}{H^2_\rf}\right) \ .
\ee
For small $\kappa_5^2 \ell U$ this reduces to the standard result
\be
B_{4D}= \frac{24\pi^2}{\kappa^4}\left(\frac{1}{\Uf}-\frac{1}{U_\M}\right) \ .
\ee
On the other hand if $\kappa_5^2 \ell U \gg 1$, brane effects will
dominate and
\be
B_{BW} = \frac{1152\pi^2}{\kappa^4\kappa_5^4\ell^2}
\left(\frac{1}{\Uf^3}-\frac{1}{U_\M^3}\right) \ll B_{4D} \ .
\ee
\\
As with CdL instanton, we see that brane effects significantly reduce
$B$, and hence increase the nucleation rate.

\section{Conclusion}

We have evaluated the bubble nucleation rate for first order phase
transitions on a Randall-Sundrum brane world. If the potential is
smaller than the brane tension, we obtain (to leading order) the
standard four dimensional results. This is not surprising, since for
small curvature, brane gravity reduces to standard four dimensional
general relativity. On the other hand if the potential is larger than
the brane tension, the non-standard brane gravity effects will be
significant. In all the cases that we considered, the nucleation rate
was significantly increased by brane gravity. This suggests that, at
least for positive definite potentials, the nucleation rate in brane
models will be higher than for conventional gravity. This is analogous
to the situation in brane cosmology, where the Hubble parameter is
larger than in the equivalent conventional cosmology.

To fully model a phase transition, we also need to consider the
expansion of the bubbles after nucleation. In other work it has been
shown that for brane models a higher nucleation rate is required for
the transitions to complete successfully~\cite{BWPT}. Our results
suggest this problem could be fixed by the brane gravity. However it
should be noted that in the early universe finite temperature
effects~\cite{Linde} will be significant, and our expressions are for
$T=0$. As well as corrections to $B$, the factor $A$ in
eq.~\eqref{Gamma} also needs to be determined. On dimensional grounds
this is usually taken to be of order $T^4$. However for $\kappa_5^2
\ell T^4 \gg 1$ brane gravity effects will be significant and we
would expect $A$ to be some combination of $T^4$ and $\kappa_5^2\ell$
instead. If this is the case the analysis of ref.~\cite{BWPT} (which
assumes $A \sim T^4$) will also be altered.


\begin{thebibliography}{99}
\bibitem{KOV}
I.~Y.~Kobzarev, L.~B.~Okun and M.~B.~Voloshin,
``Bubbles In Metastable Vacuum,''
Sov.\ J.\ Nucl.\ Phys.\  {\bf 20} (1975) 644
[Yad.\ Fiz.\  {\bf 20} (1974) 1229].
\bibitem{CC} S.~Coleman, 
``Fate of the false vacuum: Semiclassical theory,''
Phys.\ Rev.\ D {\bf 15} (1977) 2929; \\
C.~G.~Callan and S.~Coleman, 
``Fate of false vacuum. II. First quantum corrections,'' 
Phys.\ Rev.\ D {\bf 16} (1977) 1762.
\bibitem{CdL} S.~Coleman and F.~De~Luccia, 
``Gravitational effects on and of vacuum decay,''
Phys.\ Rev.\ B {\bf 21} (1980) 3305.
\bibitem{RS} L.~Randall and R.~Sundrum,
 ``An alternative to compactification,'' 
Phys.\ Rev.\ Lett.\  {\bf 83}, 4690 (1999) [hep-th/9906064].
\bibitem{BWcos} P.~Binetruy, C.~Deffayet, U.~Ellwanger and D.~Langlois,
``Brane cosmological evolution in a bulk with cosmological constant,'' 
Phys.\ Lett.\ B {\bf 477} (2000) 285 [hep-th/9910219].
\bibitem{HM} S.~W.~Hawking and I.~G.~Moss,
``Supercooled Phase Transitions In The Very Early Universe,''
Phys.\ Lett.\ B {\bf 110} (1982) 35.
\bibitem{GH} G.~W.~Gibbons and S.~W.~Hawking,
``Action Integrals And Partition Functions In Quantum Gravity,''
Phys.\ Rev.\ D {\bf 15} (1977) 2752.
\bibitem{flatO4} S.~R.~Coleman, V.~Glaser and A.~Martin,
``Action Minima Among Solutions To A Class Of Euclidean Scalar Field
Equations,'' 
Commun.\ Math.\ Phys.\ {\bf 58} (1978) 211.
\bibitem{Ida} D.~Ida,
``Brane-world cosmology,'' 
JHEP {\bf 0009} (2000) 014 [gr-qc/9912002].
\bibitem{Parke} S.~Parke,
``Gravity and the decay .of the false vacuum,'' 
Phys.\ Lett. {\bf 121B} (1983) 313.
\bibitem{BWPT} S.~C.~Davis, W.~B.~Perkins, A.~C.~Davis and I.~R.~Vernon,
``Cosmological phase transitions in a brane world,''
Phys.\ Rev.\ D {\bf 63} (2001) 083518 [hep-ph/0012223].
\bibitem{Linde} A.~D.~Linde,
``Decay Of The False Vacuum At Finite Temperature,''
Nucl.\ Phys.\ B {\bf 216} (1983) 421
[Erratum-ibid.\ B {\bf 223} (1983) 544].
\end{thebibliography}
\end{document}